\documentclass[pre,twocolumn,showpacs]{revtex4-1}

\newcommand{\Eq}[1]{Eq.~(\ref{eq:#1})}

\newcommand{\f}{\mathbf{f}}
\newcommand{\Fig}[1]{Fig.~\ref{fig:#1}}
\newcommand{\skippa}[1]{}
\newcommand{\Figure}[1]{Figure~\ref{fig:#1}}

\newcommand{\REF}[1]{Ref.~\cite{#1}}

\newcommand{\App}[1]{Appendix~\ref{sec:#1}}
\renewcommand{\f}{\mathbf{f}}
\newcommand{\rr}{\mathbf{r}}

\renewcommand{\v}{\mathbf{v}}

\newcommand{\expt}[1]{\left< #1\right>}
\newcommand{\gdot}{\dot{\gamma}}
\newcommand{\taudiss}{\tau_\mathrm{diss}}

\newcommand{\gammap}{\gamma_p}
\newcommand{\gammatau}{\gamma_\tau}
\newcommand{\etapone}{\eta_{p1}}
\usepackage{graphicx}
\usepackage{amssymb}
\usepackage{bm}

\begin{document}
\title{Considerations on the relaxation time in shear-driven jamming}

\author{Lucas Hedstr\"om}
\affiliation{Department of Physics, Ume\aa\ University, 901 87 Ume\aa, Sweden}
\author{Peter Olsson}
\email{Peter.Olsson@tp.umu.se}
\affiliation{Department of Physics, Ume\aa\ University, 901 87 Ume\aa, Sweden}

\date{\today}   

\begin{abstract}
  We study the jamming transition in a model of elastic particles under shear at zero
  temperature, with a focus on the relaxation time $\tau_1$. This relaxation time is from
  two-step simulations where the first step is the ordinary shearing simulation and the
  second step is the relaxation of the energy after stopping the shearing.  $\tau_1$ is
  determined from the final exponential decay of the energy. Such relaxations are done
  with many different starting configuration generated by a long shearing simulation in
  which the shear varible $\gamma$ slowly increases. We study the correlations of both
  $\tau_1$, determined from the decay, and the pressure, $p_1$, from the starting
  configurations as a function of the difference in $\gamma$. We find that the
  correlations of $p_1$ are more long lived than the ones of $\tau_1$ and find that the
  reason for this is that the individual $\tau_1$ is controlled both by $p_1$ of the
  starting configuration and a random contribution which depends on the relaxation path
  length---the average distance moved by the particles during the relaxation. We further
  conclude that it is $\gammatau$, determined from the correlations of $\tau_1$, which is
  the relevant one when the aim is to generate data that may be used for determining the
  critical exponent that characterizes the jamming transition.
\end{abstract}

\pacs{63.50.Lm,	
  45.70.-n	
  83.10.Rs 	
}
\maketitle

\section{Introduction}

In the everyday world there are many examples of disordered systems that change from a
flowing state to a rigid state due to a change of some control parameter. The examples are
as different as shaving cream and grains in silos. An early suggestion was the existence
of the ``jamming phase diagram'' \cite{Liu_Nagel} with the conjecture that the transition
from flowing to rigid in disordered systems has the same properties whether the control
parameter is density, shear stress, or temperature. It was however later shown that the
shear-driven jamming transition, which takes place due to the change in density at zero
temperature, is a different phenomenon than equilibrium glassy behavior \cite{Ikeda:2012,
  Olsson_Teitel:jam-T}. Another fundamental insight is that the shear-driven jamming
transition of soft particles is perfectly sharp only in the limit of vanishing shear rate
\cite{Olsson_Teitel:jamming}.

A path towards a better understanding of these systems is the study of simple models of
disorderd collections of particles through computer simulations. A complication is however
that the study of zero-temperature processes requires different methods than the ones used
in molecular dynamics.  In standard molecular dynamics the velocity of the particles
automatically makes the system explore phase space and the measurement of different kinds
of quantities along this trajectory gives ensemble averages of these quantities. For
macroscopic particles of granular matter the thermal velocity is however negligible and
ensembles of configurations need to be generated by other means. One way to achieve this
is by constantly shearing the system, commonly done with Lees-Edwards boundary conditions
\cite{Evans_Morriss}, which allow for shearing of the system indefinitely with periodic
boundaries in all directions.

The shear-driven jamming transition is characterized by the increase of shear viscosity
with increasing density and the cleanest behavior is for a system of hard disks. For hard
disks the viscosity only depends on the density, $\phi$, and the exponent $\beta$
describes the divergence of the shear viscosity at $\phi_J$,
\begin{equation}
  \label{eq:viscosity}
  \eta_\mathrm{hd}(\phi) \sim (\phi_J-\phi)^{-\beta}.
\end{equation}

A method for simulating with hard particles has been devised and was used in
\REF{Lerner-PNAS:2012}. That method is however in practice limited to rather small systems
and densities at some distance below jamming. This is so since the method relies on the
diagonalization of a matrix, which has to be done each time the contact network
changes. Another approach is to simulate soft particles at different shear rates
$\gdot$---where the limit $\gdot\to0$ corresponds to the hard disk limit---and to try to
determine the behavior in that limit by scaling analyses \cite{Olsson_Teitel:jamming,
  Olsson_Teitel:gdot-scale}. Yet another method to approach the hard disk limit---a method
of relevance for the present paper---is to start from configurations produced in the
shearing simulations, stop the shearing and let the energy relax down towards zero
energy. (Similar relaxations were first done in \REF{Hatano:2009}.) The final part of this
relaxation turns out to be exponential to an excellent approximation and each relaxation
gives a relaxation time $\tau_1$. We here use a notation where $\tau_1$, $z_1$, and $p_1$
denote quantities from single configurations or relaxations, whereas $\tau$, $z$, and $p$
denote averages over many configurations. From the configurations we also determine $z_1$
which is the average number of contacts per particle in the final configuration, after the
rattlers have been removed. (The rattlers are the particles with $\leq D$ contacts which
do not contribute to the stability of the network.) It is then found that $\tau_1$ is
directly related to the distance to isostaticity, given by $\delta z_1\equiv z_c-z_1$
\cite{Olsson:jam-relax}, with $z_c=2d$ \cite{Alexander:1998}. (See \REF{Goodrich:2012} for
the generalization of this expression to finite $N$). The relaxation time is algebraically
related to the distance to isostaticity, $\tau_1\sim \delta z_1^{-\beta/u_z}$
\cite{Olsson:jam-relax}. Here $z_c-z\sim(\phi_J-\phi)^{u_z}$ and it is commonly believed
that $u_z=1$ \cite{Heussinger_Barrat:2009}.

We also remark that it has been claimed \cite{Nishikawa_Ikeda_Berthier:2021} that the
determinations of $\tau_1$ suffer from a problematic finite size dependence which in effect
invalidates the method. As discussed in \App{FiniteSize} other studies confirm this finite
size effect but show that it is a serious problem only for certain cases and that it does
not pose a problem for the simulations as in \REF{Olsson:jam-relax}.

The determination of the relaxation time by starting from configurations obtained through
steady shearing \cite{Olsson:jam-relax} is therefore a method that gives results relevant
for the hard disk limit, that may be used for simple and clean determinations of the
critical behavior. The question however remains on how to perform such analyses
efficiently and the original motivation behind the present work was to find guidelines for
such more efficient simulations. These investigations did however lead to a number of
interesting findings both on the workings of the relaxation simulations and for the
shear-driven simulations which means that these findings are the central results whereas
the quest for an optimal approach in the simulations becomes secondary.

When approaching close to $\phi_J$ one would better make use of big system sizes both to
avoid the spurious jamming that can happen in smaller systems and to get a smaller spread
in the obtained $\tau_1$ \cite{Olsson:jam-relax}. It is also desirable to perform the
initial shearing simulations with small shear strain rates, since that somewhat speeds up
the ensuing relaxations. The question is then how to chose the distance in terms of shear
strain, $\gamma$, between successive starting configurations. The desire is to avoid
getting relaxations that are strongly correlated to each other and at the same time avoid
wasting simulation time on unnecessarily long shearing simulations between the successive
relaxations.

Beside the practical questions there are also questions on the workings of the relaxations
and one such question is the connection between the properties of the starting
configuration and the relaxed configuration. Specifically we ask to what extent it is
possible to predict $\tau_1$ from the pressure $p_1$ of the starting configuration. The
answer to this question is important both for the above stated question on the efficient
use of simulations and for a better understanding of the dynamics at densities below
jamming.

The organization of the manuscript is as follows: In Sec.~II we describe the model and the
simulations and also the scaling assumptions and the analyses employed in this work. In
Sec.~III we start by first examining the correlations of relaxation times $\tau_1$
generated from configurations a distance $\gamma$ apart. Because of the difficulty in
getting good statistics for that quantity we examine to what extent it is possible to
predict $\tau_1$ from the pressure of the starting configuration, $p_1$, and find that
$\tau_1$ is governed both by $p_1$ and by a random term which is related to the real space
distance between the relaxed configuration and its starting configuration. We then also
examine the pressure correlations and find a rich behavior, but also show that the
behavior may be understood in terms of the scaling approach. In Sec.~IV we discuss the
implications of the findings for efficient simulations with the aim of high-precision
determinations of a critical exponent. In Sec.~V we give a short summary of the
findings. We also include an appendix which discusses the possibility of logarithmic
corrections to scaling that---if present---could make the determination of the critical
divergence exceedingly difficult, and has been put forward as the reason for the
discrepancy between numerically determined exponents \cite{Olsson_Teitel:gdot-scale,
  Olsson:jam-tau, Kawasaki_Berthier:2015} and the ones obtained from a certain theoretical
approach \cite{DeGiuli:2015, H.Ikeda-logcorr:2020}.

\section{Model, simulations, and analyses}

\subsection{Simulation model}

For the simulations we follow O'Hern et al. \cite{OHern_Silbert_Liu_Nagel:2003} and
use a simple model of bi-disperse frictionless disks in two dimensions with equal numbers
of particles with two different radii in the ratio 1.4. We use Lees-Edwards boundary
conditions \cite{Evans_Morriss} to introduce a time-dependent shear strain
$\gamma = t\gdot$. We take $r_{ij}$ for the distance between the centers of two
particles and $d_{ij}$ for the sum of their radii. The relative overlap then becomes
$\delta_{ij} = 1 - r_{ij}/d_{ij}$. The interaction between two overlapping particles is
$V_p(r_{ij}) = \epsilon \delta_{ij}^2/2$; we take $\epsilon=1$. The force on particle $i$
from particle $j$ is $\f^\mathrm{el}_{ij} = -\nabla_i V_p(r_{ij})$, which means that the
magnitude becomes $f^\mathrm{el}_{ij}=\epsilon\delta_{ij}/d_{ij}$. The simulations are
performed at zero temperature. Length is measured in units of the diameter of the small
particles, $d_s$.

The total interaction force on particle $i$ is
$\f^\mathrm{el}_i = \sum_j \f^\mathrm{el}_{ij}$, where the sum extends over all particles
$j$ in contact with $i$. The simulations have been done with the RD$_0$ (reservoir
dissipation) model \cite{Vagberg_Olsson_Teitel:BagnNewt} with the dissipating force
$\f^\mathrm{dis}_i = -k_d \v_i$ where $\v_i\equiv d\rr_i/dt-y_i\gdot\hat x$ is the
non-affine velocity, i.e.\ the velocity with respect to a uniformly shearing velocity
field, $y_i\gdot\hat x$.  In the overdamped limit the equation of motion is
$\f^\mathrm{el}_i +\f^\mathrm{dis}_i = 0$ which becomes $\v_i = \f^\mathrm{el}_i/k_d$.  We
take $k_d=1$ and the time unit $\tau_0 = d_s^2 k_d/\epsilon=1$. The equations of motion
were integrated with the Heuns method with time step $\Delta t/\tau_0=0.2$.

Many quantities were measured during the shearing simulations. Two examples of relevance
for the present work are pressure $p$, and the average magnitude of the non-affine
particle velocity $v=\expt{|\v_i|}$.  We will below refer to earlier analyses of $p$ in
\REF{Olsson_Teitel:gdot-scale}. The properties of $v$ have been discussed in
\REF{Olsson:jam-vhist}; we here just note that different powers of $v$ scale differently
as criticality is approached, such that $\expt{\v_i^2}$ and $\expt{|\v_i|}^2$ diverge
differently.

\subsection{Scaling relations}

Shear-driven jamming has been found to be a critical phenomenon with shear strain rate one
of the relevant parameters, which means that jamming transition takes place at
$(\phi,\gdot)=(\phi_J,0)$ and that many properties are expected to scale with the distance
to jamming. A general review of scaling may be found in
\REF{Vagberg_Olsson_Teitel:CDn}. With $\delta\phi=\phi-\phi_J$ the pressure is expected to
scale as
\begin{equation}
  \label{eq:p-b-scale}
  p(\delta\phi,\gdot) = b^{-y/\nu}  \tilde g(\delta\phi\, b^{1/\nu}, \gdot b^z).
\end{equation}
where $\nu$ is the correlation length exponent, $z$ is the dynamical critical exponent,
and $y$ is the scaling dimension of $p$. It should be noted that this expression neglects
the correction to scaling term \cite{Olsson_Teitel:gdot-scale} which is important for
precise analyses close to jamming and has gotten a new interpretation in recent works
\cite{Olsson:jam-slow-fast-lett, Olsson:jam-slow-fast}.

With $b=\gdot^{-1/z}$ and the notation $q=y/z\nu$
this becomes
\cite{Olsson_Teitel:jamming}
\begin{equation}
  \label{eq:p.scale}
  p(\phi,\gdot) = \gdot^q g\left(\frac{\phi-\phi_J}{\gdot^{1/z\nu}}\right).
\end{equation}

To describe the deviations from the hard disk limit we note that the pressure is one of
many quantities which is $\propto\gdot$ in the $\gdot\to0$ limit. For $\eta_p\equiv
p/\gdot$ \Eq{p-b-scale} 
translates to
\begin{equation}
  \label{eq:eta-scale}
  \eta_p(\delta\phi,\gdot) = b^{z-y/\nu} \tilde h_\eta(\delta\phi\, b^{1/\nu}, \gdot b^z),
\end{equation}
and $(\phi_J-\phi) b^{1/\nu}=1$ gives $b=(\phi_J-\phi)^{-\nu}$ which in turn leads to
\begin{equation}
  \label{eq:eta.phi-scale}
  \eta_p(\phi,\gdot) = (\phi_J-\phi)^{-\beta}
  h_\eta\left(\frac{\gdot}{(\phi_J-\phi)^{z\nu}}\right),
\end{equation}
where $\beta=z\nu-y$.

The present work focuses on the relaxation time $\tau$ which, in the hard disk limit,
behaves the same as $\eta_p$ \cite{Olsson:jam-tau, Lerner-PNAS:2012},
\begin{equation}
  \label{eq:tau0}
  \tau(\phi,\gdot) = A_\tau(\phi_J-\phi)^{-\beta},\quad\gdot\to0.
\end{equation}
Another quantity of interest is the velocity per unit of shear strain which diverges as
\cite{Olsson:jam-vhist}
\begin{equation}
  \label{eq:tildev}
  \lim_{\gdot\to0} v(\phi,\gdot)/\gdot = A_v (\phi_J-\phi)^{-u_v},\quad\gdot\to0,
\end{equation}
with $u_v\approx 1.1$. For finite $\gdot$ these expressions may be generalized through the
same kind of relations,
\begin{equation}
  \label{eq:tau.scale}
  \tau(\phi,\gdot) \sim (\phi_J-\phi)^{-\beta} h_\tau\left(\frac{\gdot}{(\phi_J-\phi)^{z\nu}}\right),
\end{equation}
\begin{equation}
  \label{eq:vtilde.scale}
  v(\phi,\gdot)/\gdot \sim (\phi_J-\phi)^{-u_v}h_v\left(\frac{\gdot}{(\phi_J-\phi)^{z\nu}}\right).
\end{equation}
Note however the different nature of these quantities. The determination of the relaxation
time is from two-step simulations that are done by first shearing at a certain shear
strain rate and then stop the shearing and let the system relax to zero energy. The
non-affine velocity per shear rate, $v(\phi,\gdot)/\gdot$, characterizes the flowing
sheared steady state, whereas the relaxation time, $\tau$, is from the final stage of this
relaxation.

\subsection{Analyses}

To determine the relaxation time we run simulations as described above at zero temperature
and fixed $\gamma$ (i.e.\ with $\gdot=0$) which leads to an energy decreasing towards zero; the
simulations are aborted when the energy per particle is $E<10^{-20}$. The relaxation time
is then determined from the exponential decay of the energy per particle by fitting $E(t)$
to
\begin{equation}
  E(t) \sim e^{-t/\tau_1},\quad E(t)<10^{-17}.
  \label{eq:E-t}
\end{equation}
For each parameter set, $N$, $\phi$, and $\gdot$, the starting configurations are from a
long shearing simulation that gives a large number of different starting configurations
with different $\gamma$. The relaxation simulations then give a set of relaxation times,
$\tau_1(\gamma)$. (The individual determinations are denoted by $\tau_1$ whereas their
average is denoted by $\tau$.) On general grounds one expects two starting configurations
that differ only by a small shear $\gamma$ to be similar and to also lead to similar
relaxation times and one of the goals of the present work is to examine how quickly this
similarity decays and thus to determine the correlation shear---which is the analogous of
the correlation time---for $\tau_1$.

The autocorrelation function of some quantity $A$ is a measure of the correlations in the
fluctuations of $A$, i.e.\ $\delta A = A-\expt A$. The autocorrelation function is then
$\expt{\delta A(t')\delta A(t'+t)}$, where the average is over different initial times
$t'$.
When comparing systems that are sheared at different shear strain rates there is a trivial
shear strain rate dependence and it is therefore convenient to instead express the
correlation function in terms of the change in the shearing variable, $\gamma$,
\begin{equation}
  \label{eq:rhoA}
  \rho_A(\gamma) = \frac{\expt{\delta A(\gamma'+\gamma) \delta A(\gamma')}}{\expt{(\delta A)^2}}.
\end{equation}
We will examine this correlation function for two different quantities, $A=1/\tau_1$ and
$A=p$ and beside some results extracted from these correlation functions we will be able
to draw some conclusions about the relaxation processes.

\section{Results}

\subsection{Correlations of relaxation times}

\begin{figure}
  \includegraphics[width=7cm]{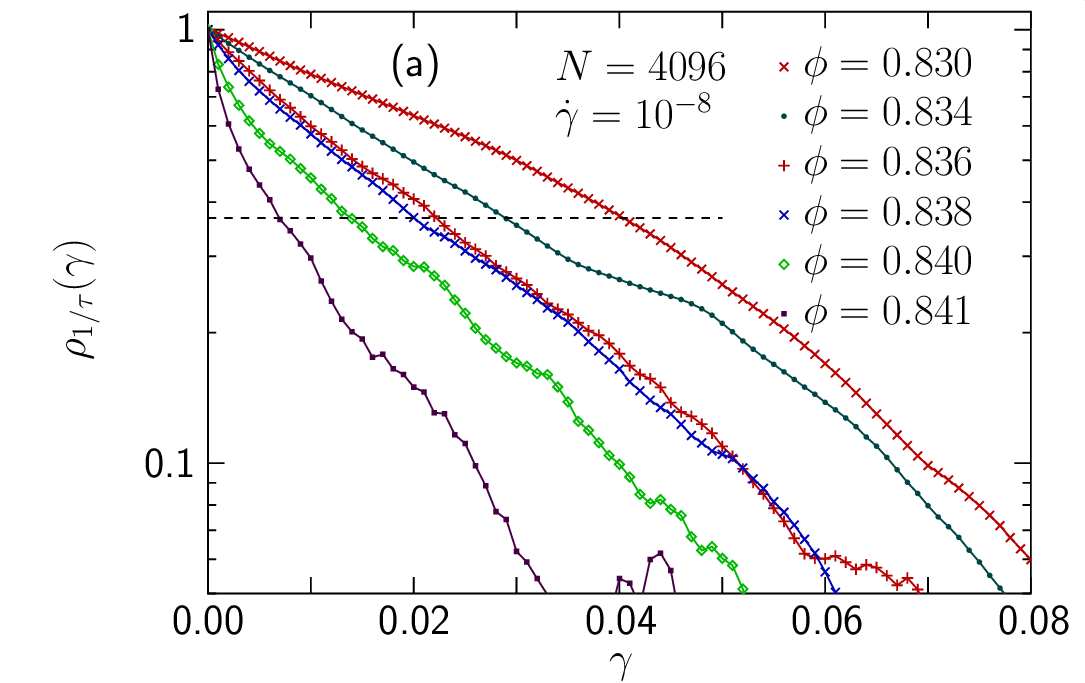}
  \includegraphics[width=7cm]{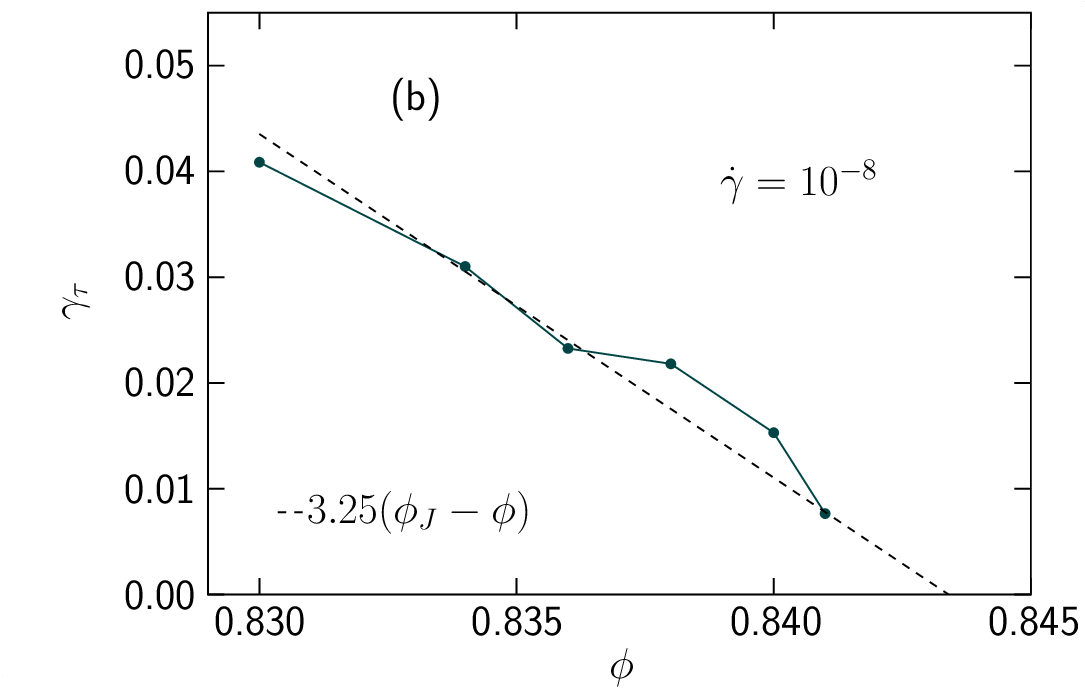}
  \includegraphics[width=7cm]{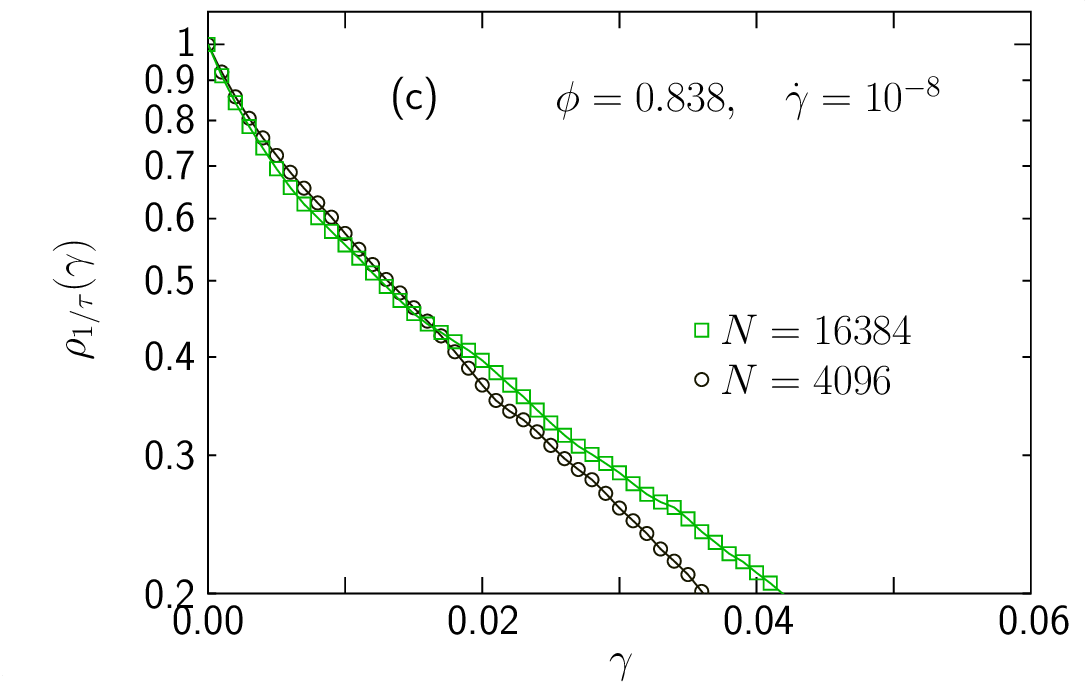}
  \caption{Correlation function for from the inverse relaxation time. Panel (a) is the
    correlation function $\rho_{1/\tau}(\gamma)$ obtained with $N=4096$ particles, shear
    strain rate $\gdot=10^{-8}$, at densities $\phi=0.830$ through 0.841. The correlation
    shear, $\gammatau$, is defined to be the value of $\gamma$ for which the correlation
    function is $\rho_{1/\tau} = e^{-1}$. Panel (b) which is $\gammatau$ vs $\phi$, shows
    that $\gammatau$ decreases as $\phi_J$ is approached from below in an approximately
    linear way, $\gammatau\sim \phi_J-\phi$. Panel (c) which shows $\rho_{1/\tau}(\gamma)$
    for two different system sizes, $N=4096$ and $N=16384$, illustrates that the
    correlation function is independent of $N$.}
  \label{fig:gamma1-rhoinvtau}
\end{figure}

To analyze correlations of $\tau_1$ we determine $\rho_{1/\tau}(\gamma)$ from \Eq{rhoA}
with $A=1/\tau_1$ for $\gdot=10^{-8}$ and $N=4096$ particles and at densities $\phi=0.830$
through 0.841, closely below $\phi_J\approx 0.8434$. The rationale for determining the
correlations of $1/\tau_1$ instead of $\tau_1$ is that $\tau_1$ can occasionally be very
big which could mean that a few big values would dominate the correlation function. This
problem is not present when instead determining the correlation of $1/\tau_1$.

\begin{figure*}
  \includegraphics[width=7cm]{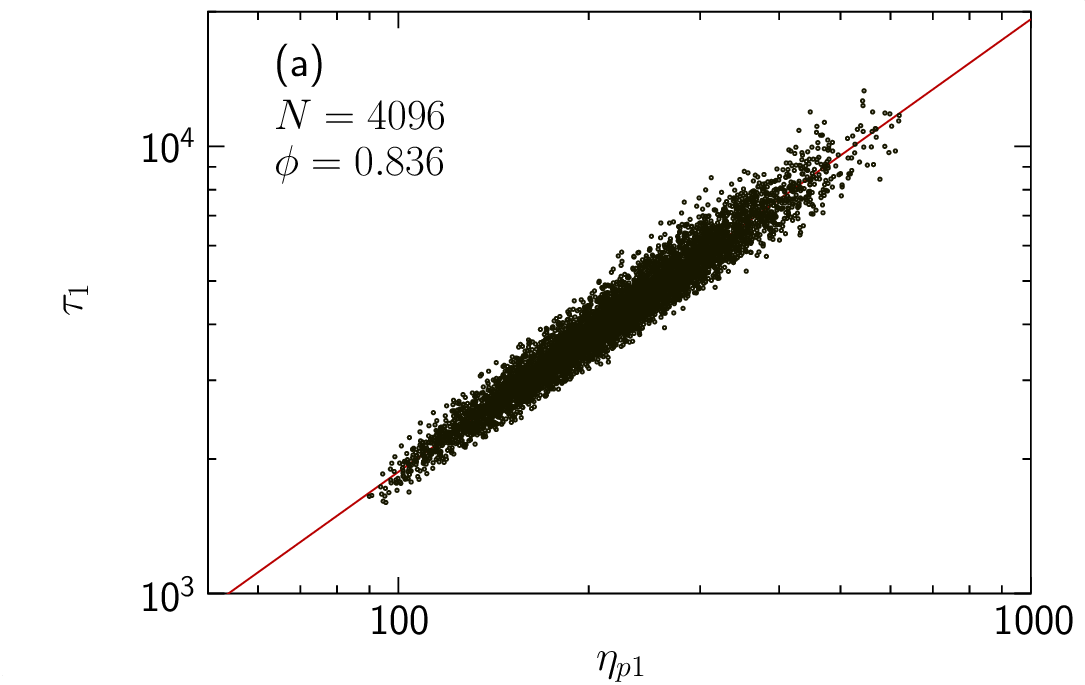}
  \includegraphics[width=7cm]{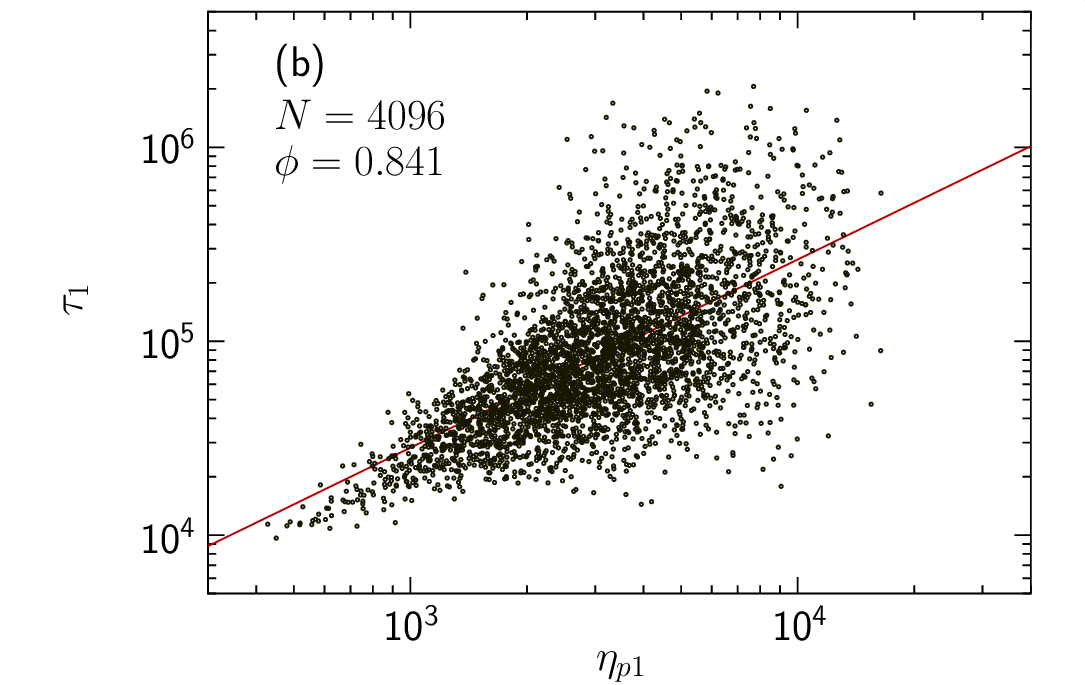}
  \includegraphics[width=7cm]{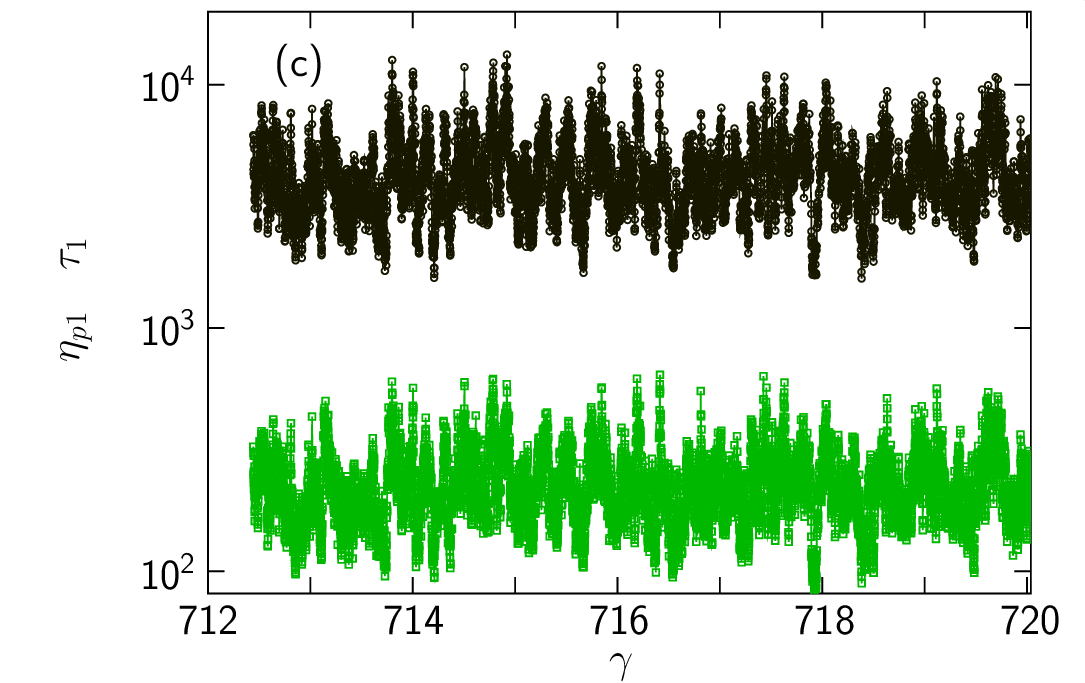}
  \includegraphics[width=7cm]{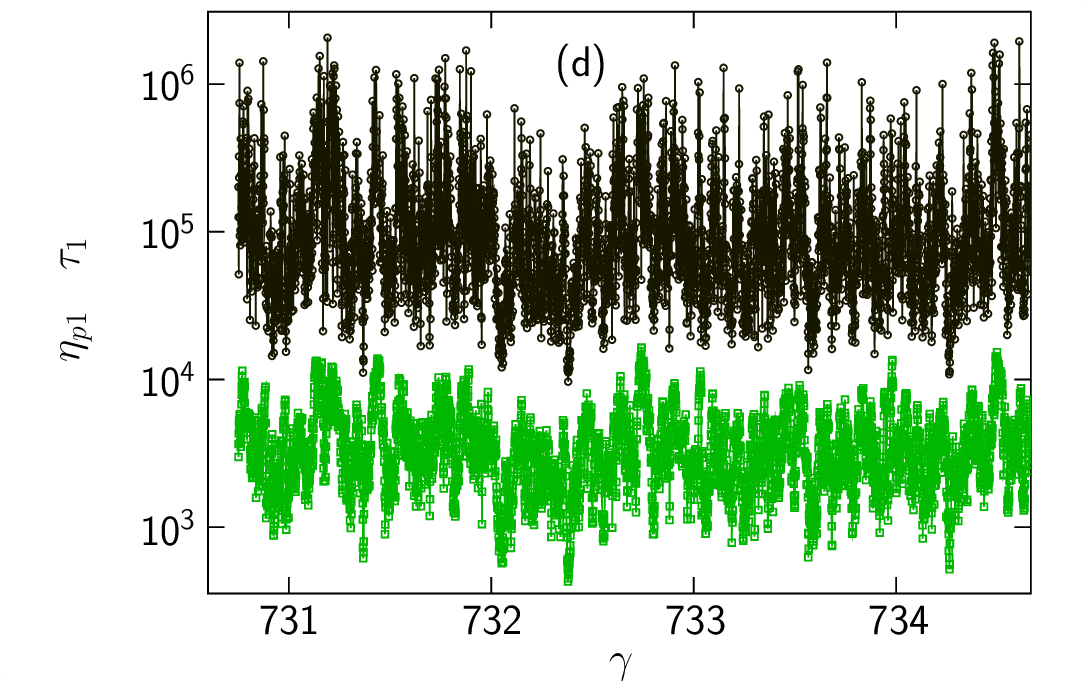}
  \caption{Plots of $\tau_1$ and $\etapone$ at two different densities, $\phi=0.836$ and
    0.841. Here $\etapone$ are from the starting configurations whereas $\tau_1$ are from
    the relaxations. The question is to what extent $\tau_1$ from the relaxation may be
    predicted from $\etapone\equiv p_1/\gdot$ of the starting configuration. The data are
    for $N=4096$ particles and shear rate $\gdot=10^{-8}$ for the shearing
    simulations. Panels (a) and be which show $\tau_1$ vs $\etapone$ shows that these
    quantities are strongly correlated for the lower density $\phi=0.836$ and more weakly
    correlated at $\phi=0.841$. The lower panels are the same data but now plotted vs
    $\gamma$ ($\sim$ simulation time).  Panel (c) is for $\phi=0.836$ whereas panel (d) is
    for $\phi=0.841$. In both cases we find that $\tau_1$ closely follows $\etapone$---the
    minima and maxima of $\etapone$ are closely followed by similar minima and maxima in
    $\tau_1$---but also that the spread is bigger in panel (d) which is for the higher
    $\phi$.}
  \label{fig:tau1-etap1}
\end{figure*}

\Fig{gamma1-rhoinvtau}(a) shows $\rho_{1/\tau}(\gamma)$. We find that each curve to a
decent approximation shows an exponential decay and that the slope of the curves increase as
$\phi$ increases towards $\phi_J$. To determine the correlation shear---the analogous
quantity to the correlation time---one should ideally fit the tail of
$\rho_{1/\tau}(\gamma)$ to an exponential decay, but since our data are rather noisy this
isn't feasible and we therefore instead determine the correlation shear $\gammatau$ from
the shear strain that gives $\rho_{1/\tau}(\gammatau)=e^{-1}$, i.e.\ from the crossings of
the horizontal dashed line in \Fig{gamma1-rhoinvtau}(a).  \Fig{gamma1-rhoinvtau}(b) shows
$\gammatau$ vs $\phi$ and we note that the figure can be taken to suggest a linear
behavior, $\gammatau\propto \phi_J-\phi$.

The data above are determined with $N=4096$ and an interesting and non-trivial question is
on the finite size dependence of $\gammatau$. One possibility would be that a larger
number of particles would give more places with important changes of the contact pattern
which would lead to a bigger change in $\tau_1$ for each change in $\gamma$.  Our results
do however suggest that there is no real finite size dependence for our quite big
systems. (For sufficiently small $N$ one would however expect changes of all kinds of
quantities.) This is illustrated by \Fig{gamma1-rhoinvtau}(c) which shows
$\rho_{1/\tau}(\gamma)$ for $N=4096$, 16384 particles, determined with $\phi=0.838$ and
$\gdot=10^{-8}$.

\Fig{gamma1-rhoinvtau}(b) is for a single shear strain rate, $\gdot=10^{-8}$, and a
further question is on the dependence of $\gamma_\tau$ on the shear strain rate.  To that
end we have attempted the same kind of analyses for different shear strain rates but found
the data to be somewhat too noisy for any safe conclusions. The basic problem is that the
determinations of $\rho_{1/\tau}(\gamma)$ requires a large number relaxation times,
$\tau_1$, which are obtained through time consuming relaxations of the system to (almost)
vanishing energy. With a limited number of such $\tau_1$ the correlation function
$\rho_{1/\tau}(\gamma)$ for different $\phi$ become quite noisy which makes it difficult
to achieve precise determinations of $\gammatau$.

As an alternative route to more insight we have therefore turned to other analyses. The
approach is based on the expectation that the individual $\tau_1$ should be strongly
influenced by properties---as e.g.\ the pressure---of the respective initial
configurations, and that the correlations between such quantities should be much easier to
determine since there are considerably more available data. For such an initial property
we here make use of the pressure, and the approach is therefore to first turn to a
comparison between relaxation time and the pressure of the corresponding initial
configuration, and then, as a second step, examine these pressure correlations. As we will
see these correlations turn out to be somewhat different and do not directly help answer
our questions. This approach does nevertheless lead to some unexpected new insights.

\subsection{Relaxation time and pressure}
\label{sec:relaxtime-and-pressure}

To examine the relation between $\tau_1$, determined from the last stage of the
relaxation, and $\etapone\equiv p_1/\gdot$, from the initial configurations (before the
relaxation), \Fig{tau1-etap1} shows $\tau_1$ and $\etapone$ for $\gdot=10^{-8}$ and
$N=4096$, at two different densities, $\phi=0.836$ and 0.841. \Fig{tau1-etap1}(a) which is
$\tau_1$ vs $\etapone$ for $\phi=0.836$ shows that these data are strongly correlated
whereas \Fig{tau1-etap1}(b), obtained at the higher density $\phi=0.841$, gives evidence
for a weaker correlation. Panels (c) and (d) show the same data but now plotted vs
$\gamma$.  For the lower $\phi$ in \Fig{tau1-etap1}(c) it is clear that the two quantities
follow each other closely but \Fig{tau1-etap1}(d), which displays the same kind of data
for the higher $\phi=0.841$, shows that the minima and the maxima of $\etapone$ are
reflected in $\tau_1$, but that there is also a big fluctuating random component to
$\tau_1$.

\begin{figure}
  \includegraphics[bb=31 322 324 660, width=4.2cm]{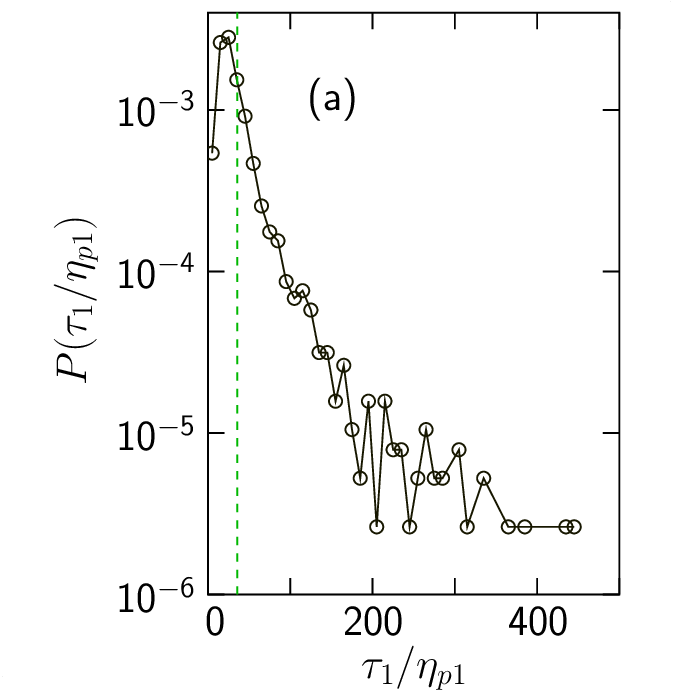}
  \includegraphics[bb=31 322 324 660, width=4.2cm]{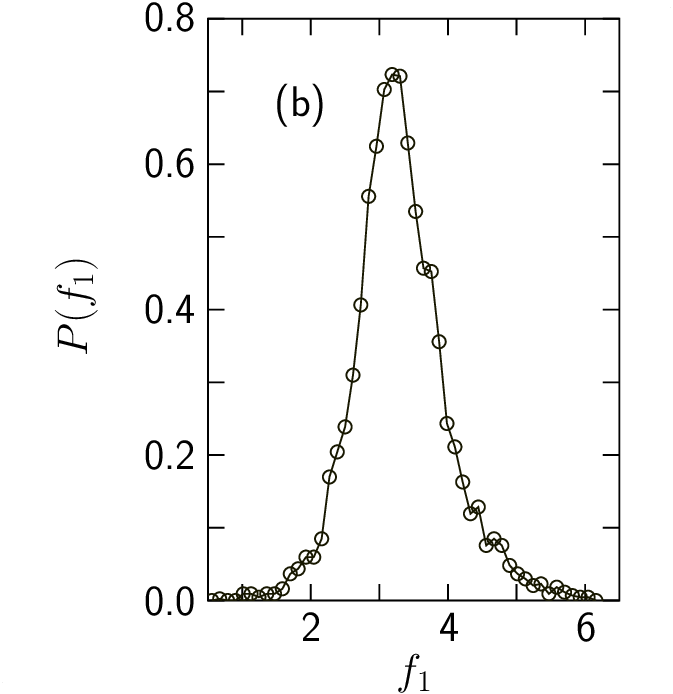}
  \includegraphics[bb=31 322 532 660, width=7cm]{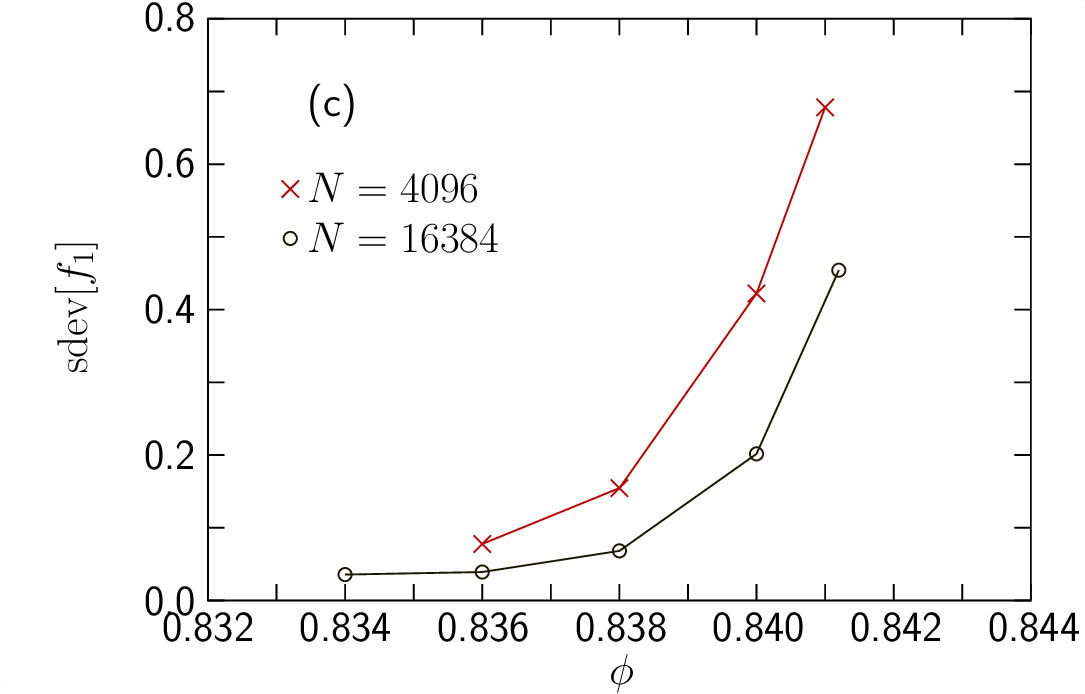}
  \includegraphics[bb=31 322 532 660, width=7cm]{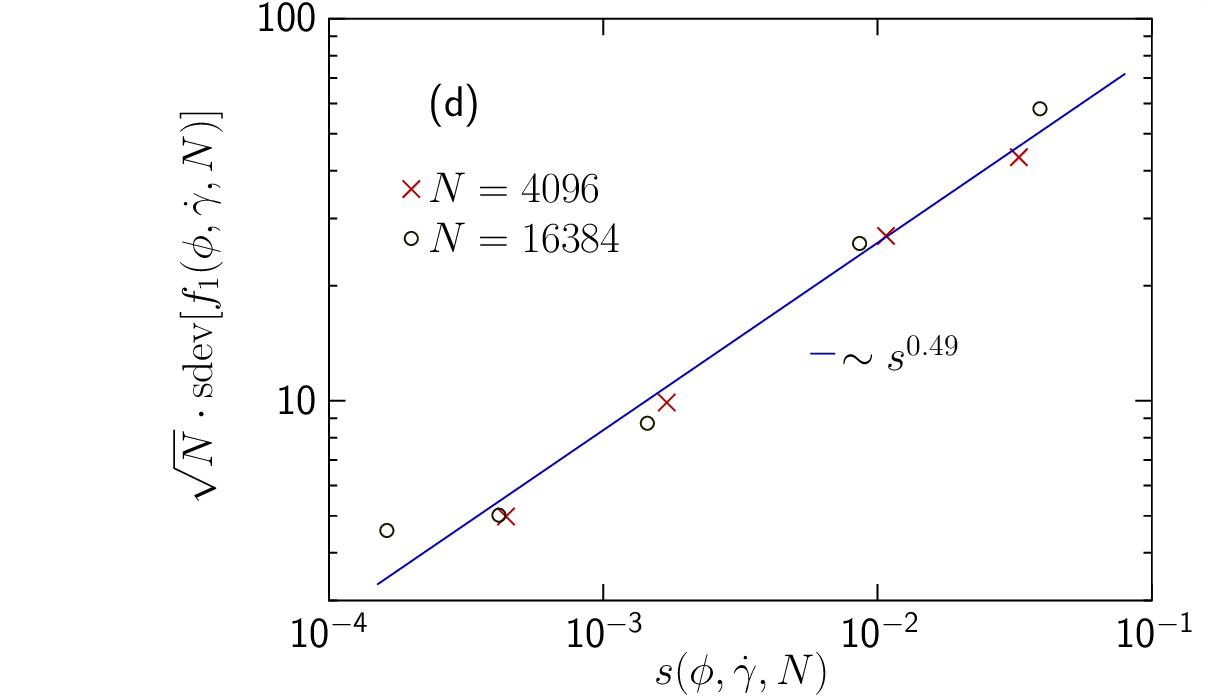}
  \caption{Properties of $f_1\equiv\ln(\tau_1/\etapone)$. All data are for
    $\gdot=10^{-8}$. Panels (a) and (b) show that the distribution of $\tau_1/\etapone$ is
    strongly skewed whereas the distribution of $f_1\equiv\ln(\tau_1/\etapone)$ is similar
    to a Gaussian distribution. This is taken to suggest that $f_1$ is a reasonable
    quantity for analyses. Panel (c) shows the spread in $f_1$---denoted by
    $\mathrm{sdev}[f_1]$---vs $\phi$ for two different sizes, $N=4096$ and 16384. The
    finite size dependence is to a good approximation given by $\sim1/\sqrt N$. Panel (d)
    shows the same data versus $s$, which is the average displacement during the
    relaxation, now compensated for the $\sqrt N$ dependence. A fit gives
    $\sqrt N\; \mathrm{sdev}[f_1]\sim s^{0.49}$, which suggests a dependence
    $\sim\sqrt s$. The conclusion is that each small $\Delta s$ from the relaxation
    contributes a random constant to $f_1$ and thus a random factor to $\tau_1/\etapone$.}
  \label{fig:sdevr-spath}
\end{figure}

To quantify this fluctuating factor we introduce
\begin{equation}
  \label{eq:f1}
  f_1 = \ln\frac{\tau_1}{\etapone},
\end{equation}
which is from the ratio of the pressure of the initial configuration and the relaxation
rate determine from relaxation simulations.  The rational for taking the logarithm is, as
shown in \Fig{sdevr-spath}(a) and (b), that the distribution of $\tau_1/\etapone$ is
strongly skewed whereas the distribution of the logarithm of the same quantity is similar
to a Gaussian distribution, as if $f_1$ were the sum of a number of independent random
variables.

If $\tau_1$ were exactly predicted by $\etapone$ $f_1$ would be a constant and to quantity
the spread in $f_1$ we determine
\begin{displaymath}
  \mathrm{sdev}[f_1]=\sqrt{\expt{f_1^2}-\expt{f_1}^2},
\end{displaymath}
which is then a measure of the size of the random fluctuating factor. \Fig{sdevr-spath}(c)
which is $\mathrm{sdev}[f_1]$ vs $\phi$ for two different sizes, $N=4096$, 16384, shows
that this quantity is small at low $\phi$ and increases rapidly with increasing
$\phi$. Further analyses (see below) suggest that $\mathrm{sdev}[f_1]\sim 1/\sqrt N$, just
as expected from elementary statistics of $N$ independent values.

The reason for the strong correlation between $\etapone$ and $\tau_1$ at the lower
densities is that the each final configuration is very close to its corresponding initial
configuration. Conversely, the bigger fluctuations of $f_1$ at higher densities suggests
the that the properties of the system often change a lot during the relaxations. As a
measure of the distance between initial and final configurations in ordinary space, we
introduce the relaxation path length which is the average distance moved by the particles
during the relaxation,
\begin{displaymath}
  s(\phi,\gdot,N) = \left<\int_0^\infty|\mathbf{v_i}(t)|dt\right>.
\end{displaymath}
The average is here over both particles and relaxation runs performed with the same
parameters, $\phi$, $\gdot$, and $N$.

\Figure{sdevr-spath}(d) is a parametric plot of $\mathrm{sdev}[f_1(\phi,\gdot,N)]$ vs
$s(\phi,\gdot,N)$ for two system sizes, $N=4096$, $16384$, shear strain rate
$\gdot=10^{-8}$, and densities $\phi=0.834$ through $0.8412$.  The system size dependence
found in \Fig{sdevr-spath}(c) is here taken care of through a factor of $\sqrt N$ and the
plotted quantity is thus $\sqrt N\cdot\mathrm{sdev}[f_1(\phi,\gdot,N)]$.  The fit to an
algebraic dependence on $s(\phi,\gdot,N)$ gives the exponent 0.49 which suggests
\begin{equation}
  \label{eq:sdevr}
  \mathrm{sdev}[f_1(\phi, \gdot, N)] \sim \frac{\sqrt{s(\phi,\gdot,N)}}{\sqrt N}.
\end{equation}
We note that this is consistent with elementary statistics if one considers $f_1$ to be
the \emph{average} of $N$ different terms which each is the sum of $\propto s$ different
terms. The logarithm in the definition of $f_1$ lets us conclude that each small
$\Delta s$ contributes a random \emph{factor} to $\tau_1$.

To summarize this part we have found that $\tau_1$ is directly controlled by $\etapone$ at
low densities and that the behavior at higher $\phi$ is similar but with big random
fluctuations. We have also found that the size of this random contribution to $\tau_1$ is
directly related to the average distance moved by the particles during the relaxation.

\subsection{Pressure correlations}

\begin{figure}
  \includegraphics[width=7cm]{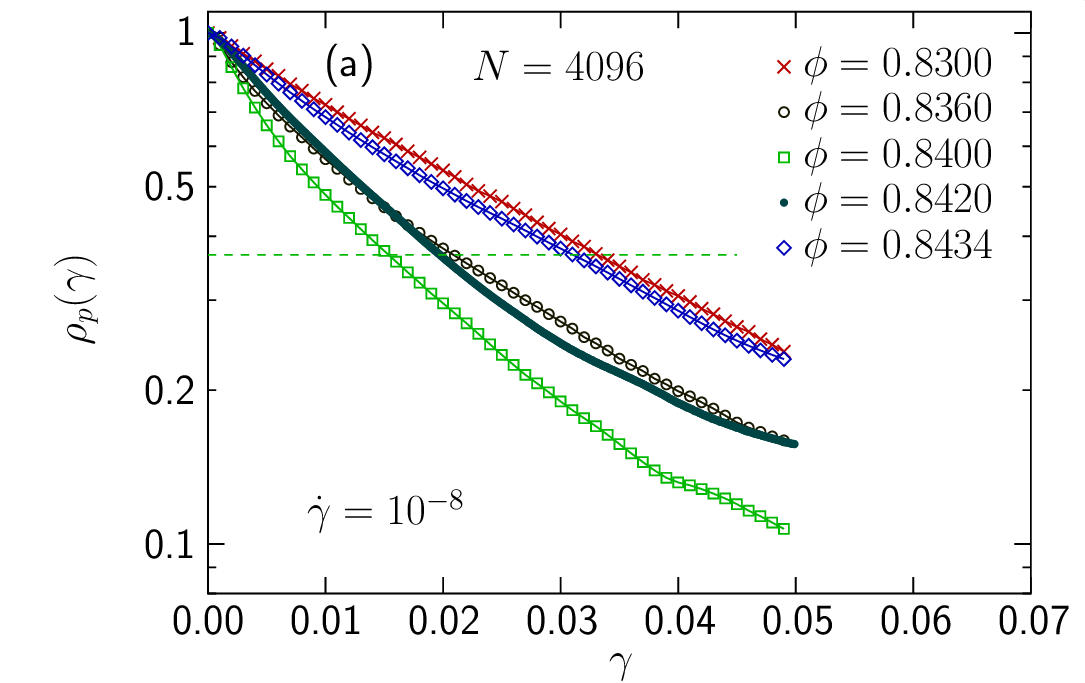}
  \includegraphics[width=7cm]{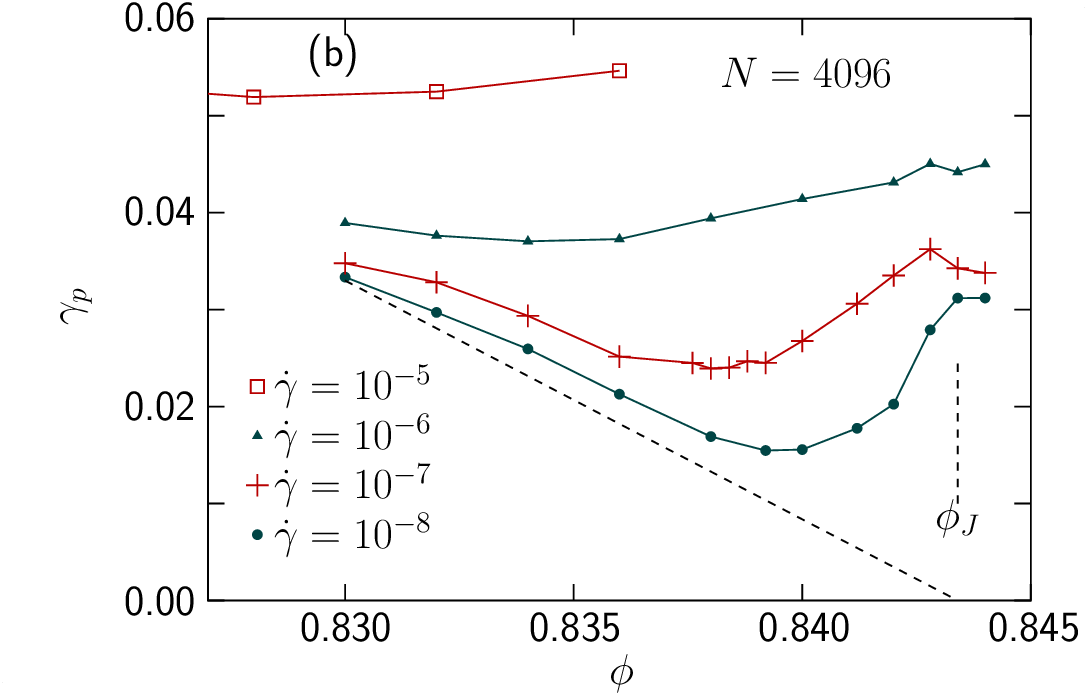}
  \includegraphics[width=7cm]{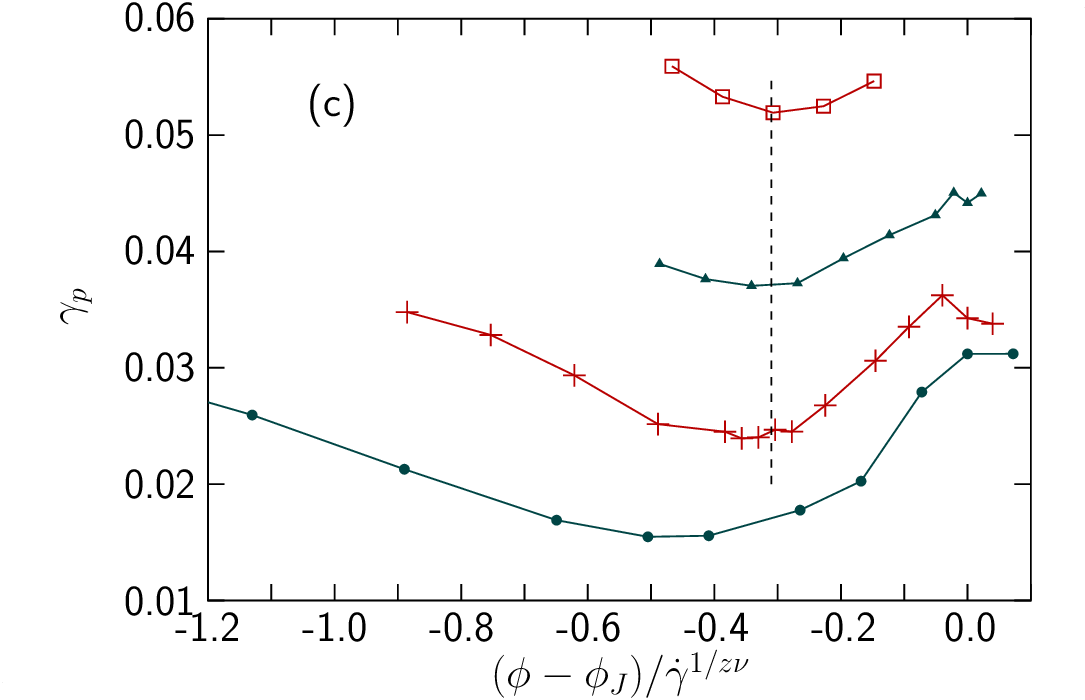}
  \caption{Correlation shear, $\gammap$, determined from $p_1(\gamma)$ measured in the
    shearing simulation. Panel (a) shows $\rho_p(\gamma)$ for $N=4096$ particles, shear
    strain rate $\gdot=10^{-8}$, and densities $\phi=0.830$ through 0.840. (Data for more
    densities are excluded in order not to clutter the figure.) The correlation shear,
    $\gammap$, is determined as the $\gamma$ for which $\rho_p(\gamma)=e^{-1}$. Panel (b)
    which is the correlation shear, $\gammap$ vs $\phi$ for three different $\gdot$, shows
    a clear dependence on $\gdot$. For each $\gdot$ there is a minimum at a certain
    $\phi_\mathrm{min}(\gdot)$. As $\gdot$ decreases the position of this minimim moves
    towards $\phi_J$. Panel (c) The same data but plotted vs the scaled distance from
    jamming as suggested by \Eq{p.scale}. We note that the minima of the respective curves
    are on top of each other, except for the data for the lowest $\gdot$. This deviation
    is tentatively attributed to a finite size effect that appears as the system size
    becomes comparable to the correlation length.}
  \label{fig:gamma1-rhop}
\end{figure}

We then turn the correlations of pressure with the aim to get an understanding of the size
of the shear strain, $\gammap$, that characterizes the decay of the pressure correlations
in shear-driven simulations.

\Figure{gamma1-rhop}(a) shows the correlation function $\rho_p(\gamma)$ for $N=4096$
particles, shear strain rate $\gdot=10^{-8}$, and densities $\phi=0.830$ through
0.8434. We find that $\rho_p(\gamma)$ decays exponentially for each $\phi$, and determine
the correlation shear $\gammap$ from the condition $\rho_p(\gammap)=e^{-1}$.
\Fig{gamma1-rhop}(b) shows $\gammap$ vs $\phi$ for different $\gdot$. The solid dots are
from the data in \Figure{gamma1-rhop}(a), the other symbols are for three higher shear
strain rates.

We note that the behavior of $\gamma_p(\phi)$ as $\gdot\to0$ suggest that $\gamma_p(\phi)$
in the hard disk limit goes approximately linearly to zero, as shown by the dashed
line. This is the same kind of behavior as shown for $\gamma_\tau$ in
\Fig{gamma1-rhoinvtau}(b) and it is also consistent with the finding that a characteristic
shear determined from the velocity-velocity correlation at $\phi_J$ vanishes as
$\gdot\to0$ \cite{Olsson:jam-vvt}. This is also directly related to the fact that
$v/\gdot$---the distance traveled per unit shear--- increases as jamming is approached.

Another interesting observation from \Fig{gamma1-rhop}(b) is that this correlation shear,
for each finite shear strain rate, $\gdot$, depends non-monotonously on $\phi$.  For each
constant shear strain rate the correlation shear, $\gammap$, first decreases towards a
minimum and then increases again when $\phi$ approaches $\phi_J$ from below. We will now
first relate this behavior to ideas from scaling, then discuss the physical mechanisms
behind this decrease and finally consider the change to an increasing trend of $\gammap$
vs $\phi$.

From the scaling assumption in \Eq{p.scale} follows that the behavior should be controlled
by the combination $(\phi-\phi_J)/\gdot^{1/z\nu}$. To test this expectation we plot
$\gammap$ vs $(\phi-\phi_J)/\gdot^{1/z\nu}$ in \Fig{gamma1-rhop}(c) and we then find that
the upward turns, to a decent approximation, take place at a constant
$(\phi_J-\phi)/\gdot^{1/z\nu}$. There is a deviation from that behavior for the smallest
shear strain rate, $\gdot=10^{-8}$, and we attribute this to a finite size effect which
could be visible when the correlation length, which increases as $\phi_J$ is approached
from below\cite{Olsson_Teitel:jam-xi-ell}, becomes comparable to the system size.

Turning to the physical reason for the decrease of $\gammap$ with increasing $\phi$, at
low $\phi$, we believe that this is an effect of the increasing particle velocity as
$\phi\to\phi_J$, described by \Eq{tildev}. The reasoning is that we expect two
configurations that are generated by the shearing dynamics should be substantially
different---such that their respective $p_1$ are also different---if the particles have on
average moved a certain characteristic distance, $\ell$. This gives the time scale
$t_v=\ell/v$ and $\gammap=t_v\gdot = \ell\gdot/v$, which together with \Eq{tildev} for the
divergence of $v/\gdot$ becomes $\gammap= (\ell/A_v) (\phi_J-\phi)^{u_v}$. The dashed line
in \Fig{gamma1-rhop}(b) is an approximate description of the behavior in the $\gdot\to0$
limit, assuming $u_v=1$. The rectilinear behavior shown there is in reasonable agreement
with the behavior described by the exponent $u_v\approx1.1$.

We now turn to the change in trend of $\gammap(\phi)$ and we are going to argue that it
goes together with a change from hard to soft particles---i.e.\ a change from negligible
to a finite particle overlaps. When the shearing is sufficiently slow that the system has
time to relax down to very small particle overlaps the system is in the hard particle
limit. Since the energy relaxation is governed by the relaxation time, $\tau$, the system
should be close to the hard disk limit as long as the relaxation time is smaller than all
other relevant time scales. With the velocity time scale, $t_v=\ell/v$, introduced above,
we get the criterion $\tau\ll t_v$, and the expectation of a change of trend when that
condition is no longer fulfilled.

We now argue that the change in behavior when $t_v\approx\tau$ is consistent with the
expectation from scaling discussed above, that the behavior should depend on the
combination $(\phi-\phi_J)/\gdot^{1/z\nu}$. For this comparison we first make use of the
relation $z\nu=\beta+y$ \cite{Olsson_Teitel:gdot-scale} to rewrite the scaling variable as
\begin{equation}
  \frac{\phi-\phi_J}{\gdot^{1/(\beta+y)}}.
  \label{eq:scaling-beta-y}
\end{equation}
From \Eq{tildev} we then find
\begin{equation}
  \label{eq:tv}
  t_v = \frac{\ell}{A_v\gdot}(\phi_J-\phi)^{u_v},
\end{equation}
which together with \Eq{tau0} and the criterion $\tau\ll t_v$ gives
\begin{equation}
  A_\tau (\phi_J-\phi)^{-\beta}\ll \frac{\ell}{A_v\gdot} (\phi_J-\phi)^{u_v},
\end{equation}
which implies that the criterion for being in the hard disk limit becomes
\begin{equation}
  \label{eq:hmm}
  \frac{\phi_J-\phi}{\gdot^{1/(\beta+u_v)}} \gg \mathrm{const}.
\end{equation}
This is similar to \Eq{scaling-beta-y} and also leads to the suggestion $u_v=y$. We also
note that the possibility that these two different exponents actually are the same is in
agreement with the numerical values in the literature, $y=1.08\pm0.03$
\cite{Olsson_Teitel:gdot-scale} and $u_v\approx1.10$ \cite{Olsson:jam-vhist}.
\begin{figure}
  \includegraphics[width=7cm]{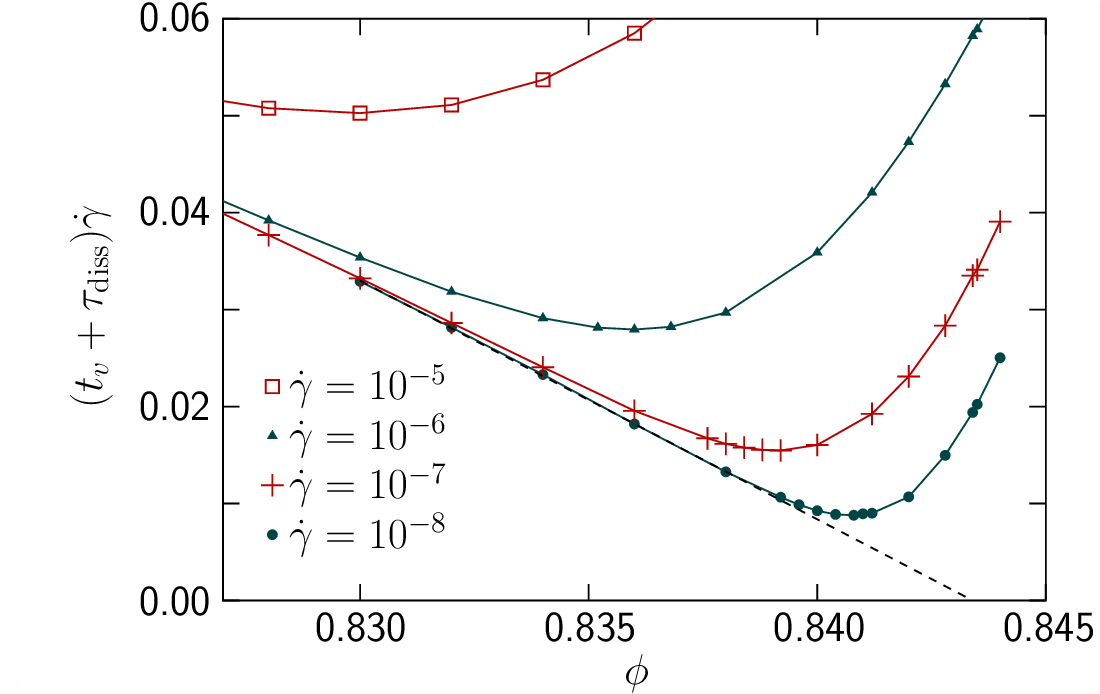}
  \caption{Attempt to predict $\gamma_p$ from our two time scales. The time scales are
    $t_v=\ell/v$ from the average particle velocity and the dissipation time $\taudiss$,
    from the initial decay of energy in a relaxation, but determined from properties
    measured in the shearing simulations and given by \Eq{taudiss}. Though the agreement
    with $\gammap$ in \Fig{gamma1-rhop}(b) is by no means perfect, we note that this
    quantity has the same general behavior.}
  \label{fig:gammap-predicted}
\end{figure}

We now also take this one step further and try to approximately express $\gammap$ in terms
of $\tau$ and $t_v$. Since $\tau$ is the time scale needed to relax energy or pressure
when shearing has been stopped, it is not unreasonable to expect $\tau$ to affect the
pressure relaxations also in the presence of shearing. However, since the conditions are
so different it follows that a possible relation between $\tau$ and the pressure
correlations could at most be an approximative one, only.

In the following, we will make use of a related but different time scale---the
\emph{dissipation time}, $\taudiss$, described in \REF{Olsson:jam-tau}. This time scale is
determined from the initial decay of a relaxation in contrast to $\tau$ which is
determined from the final part of the relaxation. These two time scales---$\tau$ and
$\taudiss$---behave the same in the $\gdot\to0$ limit but have the opposite dependence on
$\gdot$ \cite{Olsson:jam-tau}.  The decisive advantage with $\taudiss$ before $\tau$ is
however that it may be determined directly from the properties (energy and shear stress)
of the shearing simulations \cite{Olsson:jam-tau},
\begin{equation}
  \label{eq:taudiss}
  \taudiss = 2 \frac{E}{\sigma\gdot},
\end{equation}
without the need for any relaxation steps.

\Figure{gammap-predicted} shows the simplest possible way to include the effect of these
two different time scales, which is to assume that the effective correlation time is
obtained by adding together contributions from these two time scales, such that the
correlation shear is given by $\gammap = (t_v +
\taudiss)\gdot=(\ell/v+\taudiss)\gdot$. Here $v$ is the measured average velocity,
$\taudiss$ is from \Eq{taudiss} and the free parameters are $\ell=0.1$, and the prefactor
of $\taudiss$ which we take to be equal to unity. Though the agreement is by no means
perfect we note that the curves in \Figure{gammap-predicted} capture the general behavior
of the simulation data in \Figure{gamma1-rhop}(b). The conclusion is thus that the
non-monotonic behavior of $\gammap$ is an effect of the increase of $\tau$ as $\phi$
approaches $\phi_J$ from below.

\section{Discussion}

\subsection{Comparing the different correlation functions}

In the above sections we examined correlations from two different ``time'' series,
$\tau_1(\gamma)$ and $p_1(\gamma)$ and determined the related $\gammatau$ and
$\gammap$. Since the determinations of a large numbers of $\tau_1$ are quite
computationally demanding the idea was to extract the same kind of information from $p_1$
of the initial configurations which would thus reduce the need for a large number of
relaxation runs.

\Fig{gamma1-rhoinvtau} suggests that $\gammatau$ from the $\gamma$ dependence of $\tau_1$
vanishes approximately linearly as $\phi_J$ is approached from below whereas
\Fig{gamma1-rhop} shows that $\gammap$ has minima that depend on $\gdot$ and move towards
$\phi_J$ as $\gdot$ decreases. To explain the difference in behavior we then noted that
there are two contributions to $\tau_1$. The first is related to the values of $p_1$, and
thereby $\etapone$, in the initial configuration and the second is due to the relaxation
processes. The first contribution changes slowly whereas the second component fluctuates
much more rapidly and it appears that it is this second component that is responsible for
the continued decay of the correlations in \Fig{gamma1-rhoinvtau}(b) when $\gamma_p$ from
the pressure correlations in \Fig{gamma1-rhop}(c) turn upwards.

These conclusions appear to be true when $\gammatau$ and $\gammap$ are determined as the
values of $\gamma$ that make the correlation functions equal to some given constant, here
taken to be $e^{-1}$.  The correlations that are visible in the pressure correlations
ought however to be present also in $\rho_{1/\tau}(\gamma)$ and should be visible if one
were able to access the tail of $\rho_{1/\tau}$ with sufficient precision. To see this we
may consider an interval in $\gamma$ where $p_1$ is almost a constant. In this range of
$\gamma$ $\tau_1$ will vary around a certain average that depends on the average $p$. For
a different interval in $\gamma$ with another average $p$, $\tau_1$ will fluctuate around
another average value. From this consideration it appears that correlations that are seen
in $p_1$ should also be present in $\tau_1$. Since the fluctuations in $p_1$ are quite
small compared to the fluctuations in $\tau_1$ it does however seem that it would be
virtually impossible to verify the existence of these correlations in $\tau_1$.

The question is now what to conclude for efficient relaxation simulations, and what should
reasonably be the distance in terms of $\gamma$ between successive starting
configurations. It then appears that the answer---as is often the case---depends on what
the obtained data should be used for. If the goal is to get statistically independent
values $\tau_1$ for given $\phi$ and $\gdot$ it is reasonable to consider the correlations
of the initial configurations and it could then be reasonable to do the relaxation runs by
starting at points with the separation $\gamma\gg\gammap$. If, on the other hand,
the goal is to get a set of points ($\tau_1$, $\delta z_1$) for determining the exponent
$\beta/u_z$ from $\tau_1 \sim (\delta z_1)^{-\beta/u_z}$, then a possible bias of these
points towards high or small values of $\tau_1$ and $\delta z_1$ would not be a problem,
and one could then well make use of starting configurations that are quite close together
in $\gamma$, but still with $\gamma\gg\gammatau$ where $\gammatau$ is from the
apparent correlations of $1/\tau$.

\subsection{Implications for precise determinations of the critical behavior from
  relaxation simulations}

The present results together with earlier findings \cite{Olsson:jam-tau, Olsson:jam-NIB}
lead to some suggestions for determinations of points $(\tau_1, \delta z_1)$ close to
criticality for more precise determinations of the exponent $\beta/u_z$ in
$\tau\sim(\delta z)^{-\beta/u_z}$: (i) The determinations should best be done with a big
number of particles since the spread in the different quantities are $\propto1/\sqrt N$
\cite{Olsson:jam-tau}.  (ii) It is preferable to do the simulations with small $\gdot$
since one then has a lower energy to start with, but there is always a trade-off since at
large $N$ a small $\gdot$ could give very long times for generating starting
configurations with sufficient big distance in $\gamma$. [To be explicit on numbers we
note that the simulation of $10^6$ time units requires 2 hours when our parallel code is
run on 28 cores. Considering simulations at $(\phi,\gdot)=(0.842,10^{-9})$ where we have
$\gammatau(0.842)\approx 0.0045$ the simulation to advance $\gamma$ by $2\gammatau(0.842)$
would then require $\approx20$ hours.]  (iii) It would be possible to substantially speed
up the relaxation simulations by using the fast minimization protocol of \REF{Bitzek-FIRE}
instead of the slow simulations with steepest descent (which e.g.\ was used in
\REF{Olsson:jam-tau}). The final part of the simulation, which is used for the actual
determination of $\tau_1$, needs however be performed by steepest descent.  (iv) It should
be noted that the finite precision in the double precision variables that are typically
used to store the positions may lead to artifacts for very big systems
\cite{Olsson:jam-slow-fast}. This is due to two facts. First, that a larger value of a
coordinate means that fewer bits are available for storing the fractional part of the
position and, second, the fact that the \emph{net} force on a particle, which determines
the dynamics, is often considerably (i.e.\ a factor of $\tau$) smaller than the
typical \emph{ contact} force. This problem may be taken care of with a code that stores
the position in two variables, with fraction part and integer part, which ensures that
large values of the position coordinates don't affect their precision.

\section{Summary}

To summarize we have examined the correlations of both $\tau_1$ and $p_1$ as a function of
$\gamma$ motivated both by a desire to understand the basic physical mechanisms and to
answer the question on how to most efficiently determine statisticaly independent values
of the relaxation time, to be used for the determination of a critical exponent.

From $\tau_1(\gamma)$ and $p_1(\gamma)$ we determine the respective correlation shears,
$\gammatau$ and $\gammap$ and our first conclusion is on the behavior in the hard disk
limit. For the hard disk limit we conclude that our two different correlation shears both
vanish essentially linearly as $\phi\to\phi_J$ from below. We note that this is in
consistent with the earlier finding that the velocity-velocity correlation at jamming
vanishes as $\gdot\to0$ \cite{Olsson:jam-vvt}.

For $\gammap$ from the pressure correlations determined for
different $\phi$ and $\gdot$ we find that $\gamma_p(\phi)$ at constant $\gdot$ is a
non-monotonous function which first decreases with increasing $\phi$, reaches a minimum
and then increases again as $\phi\to\phi_J$. We interpret this behavior as an effect of
two different time scales where the first is directly related to the average non-affine
particle velocity, $t_v=\ell/v$, and the second is the average relaxation time,
$\tau$. Close to the hard disk limit---i.e.\ at sufficiently low $\phi$---the behavior is
dominated by $t_v$ but at higher $\phi$ the behaviour is instead dominated by $\tau$ which
diverges as $\phi\to\phi_J$.

Our data for $\gammatau$---the correlation shear for $\tau_1$---suggests that this
quantity decreases monotonously as $\phi\to\phi_J$ and we set out to analyze this
difference in behavior compared to $\gammap$. For the hard disk limit we expect
$\tau_1\propto \etapone$ \cite{Olsson:jam-tau}, and this is also borne out by our data a
low densities where the proportionality holds to a good precision for each individual
relaxation. At higher densities $\tau_1$ and $\etapone$ do however behave very differently
which is seen through $\tau_1/\etapone$ spreading considerably around its average. To
quantify this spread we determine the standard deviation in
$f_1\equiv\ln(\tau_1/\etapone)$ and find that $\mathrm{sdev}[f_1]\sim \sqrt s$, where $s$
is the average distance moved by a particle during the relaxation. Due to the logarithm in
the definition of $f_1$ we conclude that that relation implies that the contribution due
to each small $\Delta s$ is a random factor to $\tau_1$. The dependence on $N$ is in
accordance with elementary statistics of $N$ independent values, which means that this
random contribution to $\tau_1$ decreases as the system size decreases.

When it comes to efficient simulations we conclude that the distance between succcessive
configurations could reasonably be taken to be $\gamma\approx2\gammatau(\phi)$, with
$\gammatau\approx 3.25(\phi_J-\phi)$, from \Fig{gamma1-rhoinvtau}(b), which means that the
necessary distance in $\gamma$ decreases as $\phi\to\phi_J$ and that there is no need for
any very extensive shearing simulations between successive starting configurations.

\begin{acknowledgments}
  We thank S. Teitel for many comments on the manuscript. The computations were enabled by
  resources provided by the Swedish National Infrastructure for Computing (SNIC) at High
  Performance Computer Center North, partially funded by the Swedish Research Council
  through grant agreement no.\ 2018-05973.
\end{acknowledgments}

\appendix

\section{Possible problematic finite size effects}
\label{sec:FiniteSize}

We here discuss the finite size effect reported in \REF{Nishikawa_Ikeda_Berthier:2021}. In
the determinations of $\tau_1$ of \REF{Olsson:jam-relax} the starting configurations were
always from shearing simulations. It was however later argued
\cite{Ikeda__Berthier:2020-relax} that the relaxation dynamics is universal such that the
late stage of the relaxation has the same properties regardless of starting configuration
and that relaxations starting from random configurations also behave the same. That
conclusion was however based on systems of $N=3000$ particles only, and in later studies
by the same group, it was found that there is a strong finite size dependence in the
relaxation time \cite{Nishikawa_Ikeda_Berthier:2021} such that $\tau\sim\ln N$. The
suggested explanation was that a sufficiently big system will split up into different
islands with different local correlation times and that it is the biggest correlation time
that will dominate the final relaxation. With a larger number of such islands for bigger
$N$, and a simple assumption of the distribution of relaxation times, follows the
$\tau\sim\ln N$ dependence. The further conclusion was that $\tau$ is an ill-defined
quantity because of this finite-size dependence and that it may therefore not be used to
determine the critical exponent.

A later study \cite{Olsson:jam-NIB} did however modify these conclusions in several
ways. The suggested finite size dependence was confirmed, but it was also shown that the
splitting into different islands with different relaxation times only sets in for quite
big $N$, and is not the dominant mechanism for the finite size dependence. The dominant
mechanism is instead that big random initial configurations have large density
fluctuations that, to some degree, survive the relaxation process and affect the final
relaxation. This effect is not present in relaxations starting from configurations
obtained at steady shearing, since these starting configurations have a long pre-history
of a slow shearing and therefore already have an essentially uniform density. It was also
shown that it is possible to define a relaxation time which isn't plagued by the
$\ln N$-dependence, and that the problematic finite size effect is not present at the
system sizes that have been used in earlier determinations of the critical behavior
\cite{Olsson:jam-relax, Olsson:jam-3D}, and would not seem to be a problem in possible
future attempts to determine the critical behavior with higher precision.

\section{Logarithmic corrections to scaling}
\label{sec:LogCorr}

The underlying assumption in the above discussion and in the present paper is that it
should be possible to determine the critical behavior through analyses of data from
shear-driven simulations. An alternative view is that the data are affected by logarithmic
corrections that could make such analyses very difficult or perhaps altogether
unfeasible. The ground for the suggestion of logarithmic corrections to scaling is
twofold: (i) The first reason is that one could expect the presence of logarithmic
corrections in systems that are at the upper critical dimension of the model, and since
the upper critical dimension of the jamming transition is widely believed to be
$d_\mathrm{ucp}=2$ \cite{Wyart:2005, Goodrich:2012}, such corrections should be expected
in two dimensions. (ii) The second reason to consider logarithmic corrections is as an
attempt to explain the discrepancy between the values of certain critical exponents from
theoretical arguments \cite{DeGiuli:2015, H.Ikeda-logcorr:2020} and to the ones from
simulations \cite{Olsson_Teitel:gdot-scale, Olsson:jam-tau, Kawasaki_Berthier:2015}.

We do, however, not consider the evidence for the presence of logarithmic corrections to
scaling to be at all compelling. (i) A shear-driven system is in many ways different from
static jamming, which means that it could well be that the upper critical dimension is
different from two in shear-driven systems even if it were true that it is equal to two
for static jamming. (ii) A recent study \cite{Olsson:jam-slow-fast} gives reasons to
question one of the key assumptions between the theoretical approach \cite{DeGiuli:2015,
  H.Ikeda-logcorr:2020} which is that the process that governs the divergence of the
shear viscosity is ``spatially extended''. The new conclusion \cite{Olsson:jam-slow-fast} is
that the divergence of the viscosity is caused by the fastest particles
\cite{Olsson:jam-vhist} and that these particles are short range correlated, only
\cite{Olsson:jam-slow-fast}. A consequence of that reasoning is that the above mentioned
discrepancy between theory and simulations could be due to the failure of the recent
theoretical treatment \cite{DeGiuli:2015, H.Ikeda-logcorr:2020}, and that the earlier
approach of \REF{Lerner-PNAS:2012} might perhaps be correct.

We also comment on the fit of the 2D data to an expression with logarithmic correction to
scaling \cite{H.Ikeda-logcorr:2020} and argue that the fit is not as conclusive as it
could seem.  From the expectation that arguments to the logarithm function should be
dimensionless, we believe that Eq.~(12) for the relevant eigenvalue in
\REF{H.Ikeda-logcorr:2020} should rather be written with the additional free parameter
$\delta z_0$, $\lambda_1\sim(\delta z)^{\beta/u_z}|\ln(\delta z/\delta z_0)|^\alpha$
($\beta/u_z$ in our notation is their $\beta$), where both $\alpha$ and $\delta z_0$ are
unknown constants. The approach in \REF{H.Ikeda-logcorr:2020} amounts to tacitly assuming
that $\delta z_0=1$, without any reason. For $\tau\sim\lambda_1^{-1}$ the expression for
$\tau$ when including corrections to scaling becomes
\begin{equation}
  \label{eq:tau-dz-logcorr}
  \tau\sim(\delta z)^{-\beta/u_z} |\ln(\delta z) - \ln(\delta z_0)|^{-\alpha},
\end{equation}
and it is then evident that the value of $\delta z_0$ can not be absorbed in a prefactor.
The fact that the value of the exponent, $\beta/u_z=3.41$, which fits the data in 3D
without corrections to scaling, happens to give a good fit of the 2D data to
\Eq{tau-dz-logcorr} with $\delta z_0=1$, then appears to be fortuitous only, since there
is no reason to believe that $\delta z_0=1$ actually is the correct value. When taking
both $\alpha$ and $\beta/u_z$ to be free parameters one expects to find that acceptable
fits are obtained with a wide range of values of $\beta/u_z$. This should be kept in mind
when trying to assess the evidence of \REF{H.Ikeda-logcorr:2020}. It also seems that
ordinary corrections to scaling could give an equally good fit.

We finally comment on a possible and reasonable criticism of the determinations of the
exponents in \REF{Olsson_Teitel:gdot-scale}, which is that they should be regarded with
quite some skepticism since the resulting exponents were obtained through a fitting of
data with not too much structure to the sum of two unknown scaling functions---a main term
and a correction term---which are both from polynomials with quite a number of fitting
parameters. It could then seem that this complicated analysis may well give exponents that
are somewhat off the correct values even without possible additional complications as the
need for logarithmic corrections to scalinig. This concern is however in effect addressed
in \REF{Olsson:jam-slow-fast} since the origin of the correction term is there identified
which \emph{makes the correction term known} apart from a multiplicative factor. The
correction term is there found to be given by $\mathrm{const}\times W_p(\phi,\gdot)$,
where $W_p(\phi,\gdot)$ is determined from the velocity distribution. This means a great
simplification of the analysis and the fact that that approach gave similar values of the
critical exponents to the ones in \REF{Olsson_Teitel:gdot-scale}, gives reason to believe
that these earlier results actually were correct.

%

\end{document}